\documentclass[12pt]{article}
\usepackage{scicite}

\usepackage{times}
\usepackage{amsmath}
\usepackage{amssymb}
\usepackage{color}
\usepackage{graphicx}
\usepackage{lineno}

\newenvironment{sciabstract}{%
	\begin{quote} \bf}
	{\end{quote}}

\newcommand{\ket}[1]{\left |#1\right\rangle}

\title{Quantum optical circulator controlled by a single chirally coupled atom}

\author{Michael Scheucher,$^{1}$, Ad\`ele Hilico,$^1$, Elisa Will,$^1$, J\"urgen Volz,$^{1\ast}$ \& Arno Rauschenbeutel$^{1\ast}$\\
	\\
	\normalsize{$^{1}$Vienna Center for Quantum Science and Technology, TU Wien -- Atominstitut,}\\
	\normalsize{Stadionallee 2, 1020 Vienna, Austria}\\
	\\
	\normalsize{$^\ast$To whom correspondence should be addressed; E-mail:  jvolz@ati.ac.at, arno.rauschenbeutel@ati.ac.at.}
}

\begin{document}

\baselineskip24pt

\maketitle 

\begin{sciabstract}
We demonstrate a fiber-integrated quantum optical circulator that is operated by a single atom and that relies on the chiral interaction between emitters and transversally confined light. Like its counterparts in classical optics, our circulator exhibits an inherent asymmetry between light propagation in the forward and the backward direction. However, rather than a magnetic field or a temporal modulation, it is the internal quantum state of the atom that controls the operation direction of the circulator. This working principle is compatible with preparing the circulator in a coherent superposition of its operational states. Such a quantum circulator may thus become a key element for routing and processing quantum information in scalable integrated optical circuits. Moreover, it features a strongly nonlinear response at the single-photon level, thereby enabling, e.g., photon number-dependent routing and novel quantum simulation protocols.
\end{sciabstract}

In the same way as their electronic counterparts, integrated optical circuits require nonreciprocal elements, like diodes and circulators, for signal routing and processing. Bulk optical implementations of such components are readily available and rely mostly on nonreciprocal polarization rotation via the Faraday effect~\cite{Meschede2007}. However, this mechanism cannot straightforwardly be translated to integrated optics because nano-optical structures are typically birefringent. Demonstrations of integrated nonreciprocal devices therefore rather employed, e.g., nonlinear optical effects~\cite{Gallo01, Fan12b, Peng14}, time-modulation of the waveguide~\cite{Lira12, Tzuang2014, Kim15}, or magneto-optical effects in conjunction with the extraordinary polarization properties of strongly confined light fields~\cite{Shoji08, Tien11, Bi11, Shoji14}.
Still, none of these approaches could, so far, simultaneously realize strong nonreciprocity, low loss, and compatibility with low light levels. However, these characteristics are crucial when it comes to quantum applications, like quantum communication~\cite{Gisin2007}, quantum information processing~\cite{Nielsen2011}, and quantum simulation~\cite{Feynman1982}. There, information is encoded in individual photons and their loss must be avoided as much as possible. This condition narrows down the scope of quantum-compatible nonreciprocal optical elements to nonreciprocal phase shifters and circulators.

\begin{figure}[tb]
	\centering
	\includegraphics[width=14cm]{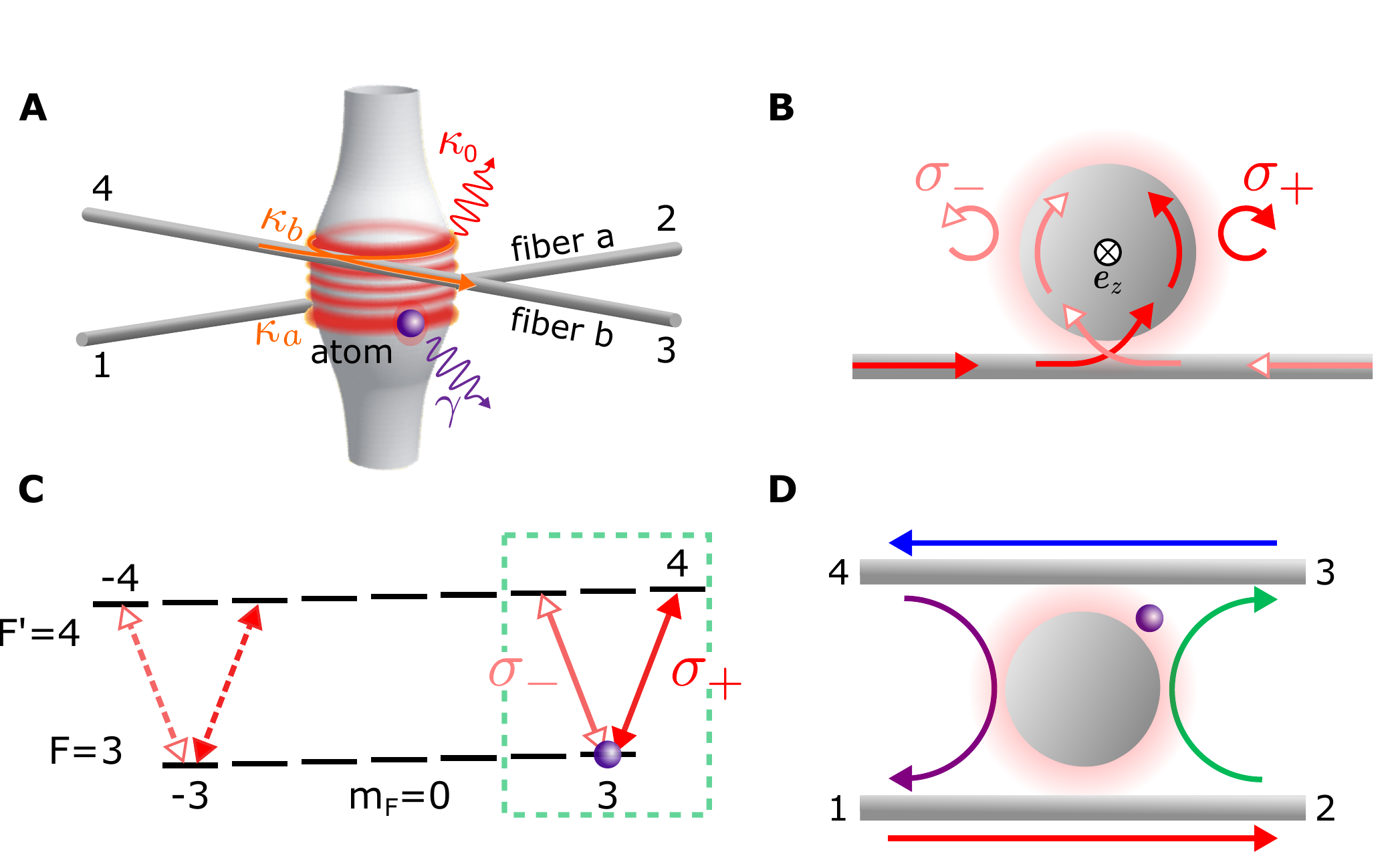}
	\caption{\textbf{Operation principle of the circulator.} (\textbf{A}) Schematic of the experimental system. A single rubidium 85 atom is coupled to the whispering-gallery-mode of a bottle microresonator which is interfaced by two tapered fiber couplers with subwavelength-diameter waists.  (\textbf{B}) The polarization of the evanescent field of the modes, $\sigma-$ and $\sigma^+$, depends on their propagation direction, clockwise (cw) or counterclockwise (ccw) and, hence, on the direction in which the light is launched through the coupling fiber.  (\textbf{C}) When a $^{85}$Rb atom is prepared in the $F=3$, $m_F=+3$ Zeeman state, the transition strength for $\sigma^+$ polarized light (i.e., for the ccw mode) is 28 times larger than for $\sigma^-$ polarized light (i.e., for the cw mode). This situation is reversed if the atom is prepared in $m_F=-3$, where $\sigma^-$ polarized light couples much more strongly to the atom.  (\textbf{D}) Routing behaviour of the circulator for the atom prepared in $m_F=+3$. Light is redirected from input port 1 to output port 2, port 2 to port 3, 3 to 4, and 4 to 1.}
	\label{fig:chiral_light-matter-interaction_and_circulator}
\end{figure}

Here, we experimentally realize a fiber-integrated circulator that is capable of routing individual photons for quantum applications. It is operated by a single atom that is coupled to the evanescent field of a whispering-gallery-mode (WGM) microresonator which is interfaced with two coupling fibers~\cite{Poellinger2010,OShea2013}, thereby realizing a four-port device, see Fig.~\ref{fig:chiral_light-matter-interaction_and_circulator}A. 
Most distinctively, it is the internal quantum state of the atom that controls the operation direction of the circulator. Thus, the circulator can in principle be prepared in a coherent superposition of its operational states, thereby turning it into an intrinsically quantum device. The corresponding nonreciprocal unitary quantum operation provided by these devices may become a key ingredient for processing quantum information in scalable integrated optical circuits. Moreover, being operated by a single atom, our circulator features a strongly nonlinear response at the single-photon level. We demonstrate that this allows us to perform photon number-dependent routing which has applications in, e.g., novel quantum simulation protocols.

In the following, we describe the operation principle of the circulator: In order to achieve efficient routing, the coupling rates, $\kappa_a$ and $\kappa_b$, between the resonator field and the field in the respective coupling fiber, $a$ or $b$, are adjusted such that both fibers are approximately critically coupled to the empty resonator, i.e., $\kappa_a\approx\kappa_b\gg\kappa_0$, where $\kappa_0$ is the intrinsic resonator field decay rate. When no atom is coupled to the resonator mode, this then realizes an add--drop filter~\cite{Poellinger2010,OShea2013} where light that is launched into one fiber will be transferred to the other fiber via the resonator. Due to its strong transverse confinement, the evanescent field of the clockwise (cw) propagating resonator mode is almost fully circularly polarized~\cite{Junge13}. Its electric field vector rotates counterclockwise in the plane orthogonal to the resonator axis ($z$-axis), corresponding to $\sigma^-$-polarization, see Fig.~\ref{fig:chiral_light-matter-interaction_and_circulator}B. Time-reversal symmetry then implies that the evanescent field of the counterclockwise (ccw) propagating mode is almost fully $\sigma^+$-polarized~\cite{Junge13}. Coupling these `spin--orbit-locked' evanescent fields of WGMs to a single atom recently enabled the implementation of an optical switch that is controlled by a single photon~\cite{Shomroni2014}. Moreover, an optical isolator was realized which either transmits or dissipates fiber-guided light depending on its propagation direction~\cite{Sayrin2015bb}. For the circulator, we resonantly couple a single $^{85}$Rb atom to the resonator. It is prepared in the outermost Zeeman sublevel $m_F=+3$ of the $5S_{1/2}$, $F=3$ hyperfine ground state and, thus, the counter-propagating resonator modes couple to an effective $V$-level system, see Fig.~\ref{fig:chiral_light-matter-interaction_and_circulator}C. Importantly, the strength of the transition to the $5P_{3/2}$, $F'=4$, $m_{F'}=+4$ excited state is 28 times stronger than that to the $F'=4$, $m_{F'}=+2$ state~\cite{metcalf1999}. As a consequence, light in the ccw mode interacts strongly with the atom with a coupling strength $g_{\rm ccw}$. In contrast, light in the cw mode exhibits much weaker coupling $g_{\rm cw} \ll g_{\rm ccw}$. This chiral light--matter interaction breaks Lorentz reciprocity~\cite{Jalas13, Xia14,Lenferink14,Sayrin2015bb}, and the presence of the atom changes the resonator field decay rate from $\kappa_{\rm tot}=\kappa_0+\kappa_a+\kappa_b$ to $\kappa_{\rm tot}+\Gamma_{{\rm cw}/{\rm ccw}}$, where $\Gamma_{{\rm cw}/{\rm ccw}}=g_{{\rm cw}/{\rm ccw}}^2/\gamma$ is the direction-dependent atom-induced loss rate~\cite{supplementaries} and $\gamma=2\pi\times 3$ MHz is the dipole decay rate of Rb. For light in the cw mode, $\Gamma_{\rm cw}$ is small and the field decay rate is not significantly modified by the atom, while for the ccw mode, $\Gamma_{\rm ccw}$  can become comparable to or even exceed $\kappa_{\rm tot}$. Consequently, the add--drop functionality is maintained when light is launched into those fiber ports for which it couples to the cw mode, i.e., input ports 2 and 4 in Fig.~\ref{fig:chiral_light-matter-interaction_and_circulator}D. However, for the two other input ports (1 and 3 in Fig.~\ref{fig:chiral_light-matter-interaction_and_circulator}D), the light couples to the ccw mode and the resonator--atom system operates in the undercoupled regime, 
\begin{equation}
\kappa_{a},\kappa_b\ll\Gamma_{\rm ccw}.\label{eqn1}
\end{equation}
In this case, the incident light field remains in its initial fiber. Overall, the device thus realizes an optical circulator that routes light from the input port $i$ to the adjacent output port $i+1$ with $i\in \{1,2,3,4\}$, see Fig.~\ref{fig:chiral_light-matter-interaction_and_circulator}D. Preparing the atom in the opposite Zeeman ground state, $F=3$, $m_F=-3$, exchanges the roles of the cw and ccw mode and thus yields a circulator with reversed operation direction. Hence, the circulator is programmable and its operation direction is defined by the internal state of the atom. 

For near perfect circular polarization of the modes and our experimental parameters ($g_\text{ccw}\approx2\pi\times12$~MHz), the ratio between $\Gamma_{\rm ccw}\approx2\pi\times48$ MHz and $\Gamma_{\rm cw}\approx2\pi\times1.7$~MHz is finite. Concerning the performance of the circulator, there is, thus, a trade-off between efficient light transfer from one fiber to the other via the cw mode, which implies $\kappa_a,\kappa_b \gg \kappa_0+\Gamma_{\rm cw}$, and the condition that the presence of the atom should significantly influence the field decay rate, see equation~(\ref{eqn1}). To find the optimum working point in our experiment, we measure the circulator performance as a function of the fiber--resonator coupling strengths, $\kappa_a$ and $\kappa_b$, which can be adjusted by changing the distance between the respective fiber and the resonator surface. We impose the constraint that fiber $a$ is critically coupled to the empty resonator which is loaded with fiber $b$: $\kappa_a=\kappa_b+\kappa_0$~\cite{supplementaries}. For each setting, we measure the transmissions $T_{i,j}$ to all output ports $j$  when sending a weak coherent probe field into the four different input ports $i$~\cite{supplementaries}. 
Figures~\ref{fig:transmission_scan}A and B show the relevant transmissions as a function of $\kappa_{\rm tot}$, where the solid lines are the theoretical prediction for our system~\cite{supplementaries}.

To evaluate the performance of the circulator, we compare the measured transmission matrix $(T_{i,j})$ to the transmission matrix  $(T_{i,j}^{\rm id})$ expected from the ideal circulator which is given in Fig~\ref{fig:transmission_matrix}A. In order to quantify the overlap with the ideal device, we define the operation fidelity of our circulator by
\begin{equation}
\mathcal{F}=1-\frac{1}{8}\sum_{i,j}|\frac{T_{i,j}}{\eta_i}-T_{i,j}^{\rm id}|  \;.
\end{equation}
Here, $\eta_i=\sum_k T_{i,k}$ is the survival probability of a photon entering port $i$, i.e., the probability that the photon is recovered at any of the four output ports. The minimum fidelity is $\mathcal{F}=0$, while $\mathcal{F}=1$ is reached for an ideal operation. For any reciprocal device ($T_{i,j}=T_{j,i}$) the fidelity is bound by  $\mathcal{F}\leq0.5$. In Fig.~\ref{fig:transmission_scan}C and D, we plot $\mathcal{F}$ and the average photon survival probability $\eta=\sum_i \eta_i/4$ as a function of $\kappa_{\rm tot}$. The results show an optimum circulator performance for $\kappa_{\rm tot}/2\kappa_0=2.2$, where $\mathcal{F}=0.72\pm0.03$ and, at the same time, $\eta=0.73\pm 0.04$. The transmission matrix $(T_{i,j})$ for the optimum working point is plotted in Fig.~\ref{fig:transmission_matrix}B and shows good qualitative agreement with that of an ideal circulator, see Fig.~\ref{fig:transmission_matrix}A. In order to demonstrate that the chiral atom--light coupling is at the origin of the nonreciprocal behavior, we also measure the transmission matrix without coupled atom and obtain a symmetric matrix, see Fig.~\ref{fig:transmission_matrix}D. The circulator performance can also be quantified by the isolations, $I_{i}=10\log( T_{i,i+1}/T_{i+1,i})$, of the four optical diodes formed between adjacent ports. For the optimum working point, we obtain  $(I_{i})=(10.9 \pm 2.5,6.8 \pm 1.3,4.7\pm 0.7,5.4\pm 1.1)$ dB and an average insertion loss of $-10\log{\eta}=1.4$~dB.

In order to reverse the operation direction of the circulator, we now prepare the atom in the opposite Zeeman ground state, $F=3$, $m_F=-3$~\cite{supplementaries}. This results in a complementary $V$-type level scheme, see dashed arrows in Fig.~\ref{fig:chiral_light-matter-interaction_and_circulator}C. For this case, we obtain the transmission matrix shown in Fig.~\ref{fig:transmission_matrix}C, again measured for the optimum fiber--resonator coupling rate. Here, we observe a fidelity with respect to the reversed circulator of $\mathcal{F}=0.70 \pm0.02$, a photon survival probability of $\eta=0.69\pm 0.02$, and optical isolations $(I_{i})=-(8.3 \pm 0.8, 4.9 \pm 0.7, 3.7 \pm 0.4, 5.6 \pm 0.5)$~dB. Taking into account the sign change, these results agree well with the values obtained for the atom in the $F=3$, $m_F=+3$ state.

\begin{figure}[t]
	\centering
	\includegraphics[width=12cm]{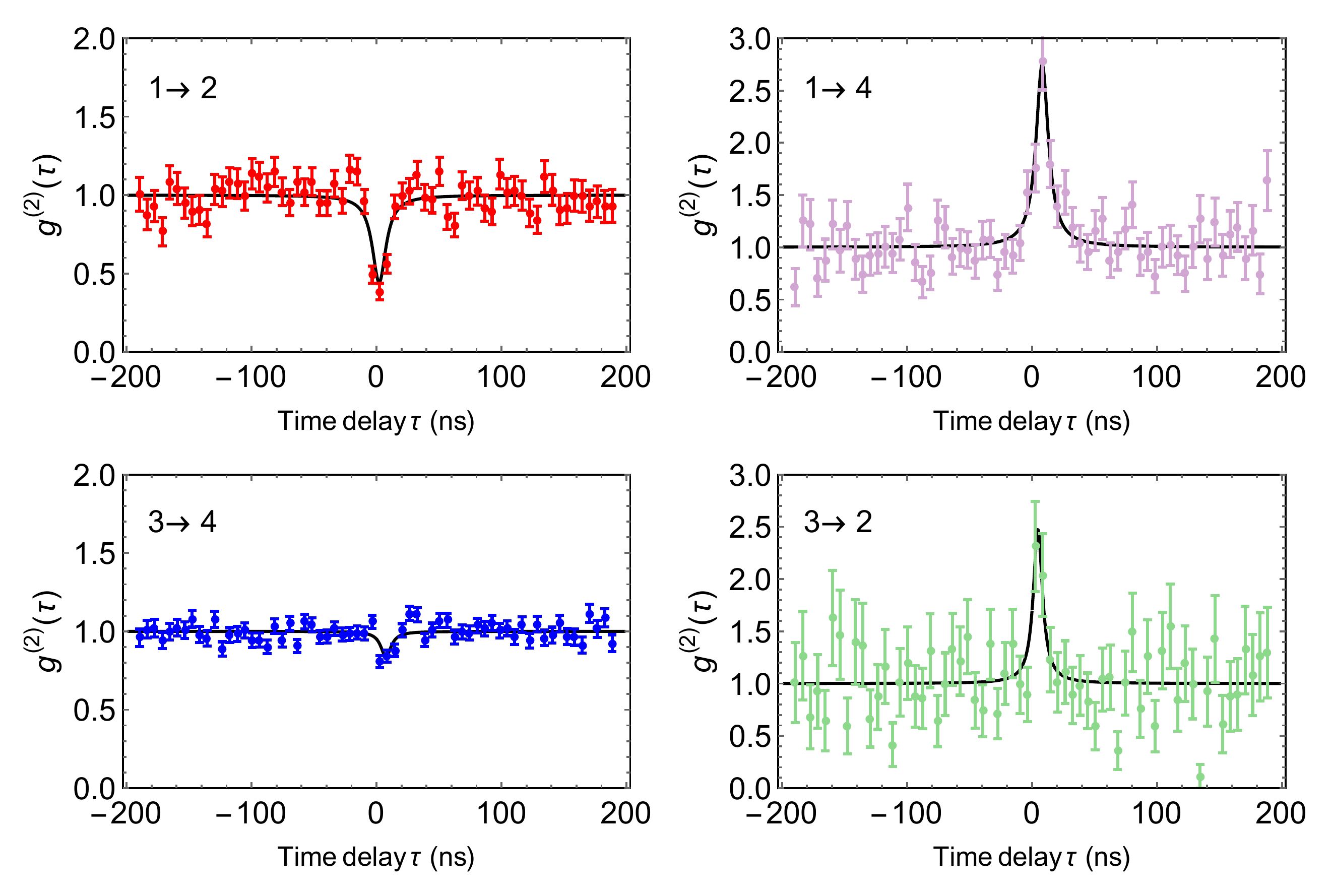}
	\caption{\textbf{Quantum nonlinearity of the circulator.} Second-order correlation, $g^{(2)}(\tau)$, as a function of the detection time delay $\tau$ between pairs of photons, normalized such that $g^{(2)}(\tau)=1$ for $\tau\gg1/\kappa_{\rm tot}$. The labels $i \to j$ indicate the input and output ports for the respective measurement. The solid lines are guides to the eye. The error bars indicate the $\pm 1\sigma$ statistical error. For the settings $1 \to 2$ and $3 \to 4$, we observe photon antibunching which is more pronounced in the former case due to the unequal fiber couplings, $\kappa_a > \kappa_b$. Photon bunching of similar amplitude is apparent for the settings $1 \to 4$ and $3 \to 2$. This agrees with the theoretical expectation because, here, both measurements amount to probing the photon statistics of the intra-resonator field. We note that the fact that $g^{(2)}(0)$ does not reach zero for the settings $1 \to 2$ and $3 \to 4$ is not due to experimental imperfections but is theoretically expected as the light fields at output ports $2$ and $4$ are in a coherent superposition between the (bunched) resonator light field that couples back into the fiber and the stronger coherent state of the probe laser field.}
	\label{fig:twophoton}
\end{figure}

Finally, we expect a strongly nonlinear optical response of the circulator down to the level of single photons since, in the regime of strong coupling, a single photon already saturates the atom~\cite{Shomroni2014,Volz2014,OShea2013}. In particular, the transmission properties for the case of two photons simultaneously impinging on the circulator should strongly differ from the single-photon case. In order to demonstrate this quantum nonlinearity, we measure second-order correlation functions for all input--output configurations when the atom is prepared in $F=3$, $m_F=+3$. Figure~\ref{fig:twophoton} shows exemplarily those second-order correlation functions which yield the strongest signals. As expected, they occur for the cases where the photons couple into the ccw resonator mode. We observe photon antibunching when the photons remain in the initial fiber (forward direction of the circulator: $1\rightarrow 2$, $3\rightarrow 4$). When the photons are transmitted to the other fiber (backward direction of the circulator: $1\rightarrow4$, $3\rightarrow2$), we observe clear photon bunching. This behaviour illustrates the photon number-dependent routing capability: while individually arriving photons remain in their original fiber, simultaneously arriving photons are preferentially transferred to the other fiber.

The demonstrated circulator concept is per se useful for the processing and routing of classical signals at ultra-low light levels in integrated optical circuits and networks. Beyond that, and in contrast to dissipative nonreciprocal devices, a circulator that is controlled by a single quantum system also enables operation in coherent superposition states of routing light in one and the other direction, thereby paving the way towards its application in future photonic quantum protocols. The demonstrated operation principle is universal in the sense that it can be straightforwardly implemented with a large variety of different quantum emitters provided that they exhibit circularly polarized optical transitions and that they can be spin-polarized. Using state-of-the-art WGM microresonators \cite{Poellinger2009}, one could realize a circulator with optical losses below 6\% and close-to-unit operation fidelity~\cite{supplementaries}. This would then allow one to almost deterministically process and control photons in an integrated optical environment. Moreover, networks of such quantum circulators are potential candidates for implementing lattice-based quantum computation~\cite{Raussendorf2001}. And finally, such networks would allow one to implement artificial gauge fields for photons~\cite{Koch2010,Hafezi2012,Schmidt2015} where a nonlinearity at the level of single quanta allows for the flux to become a dynamical degree of freedom that interacts with the particles themselves~\cite{Walter2015}.


\section*{Acknowledgements}
\noindent The authors are grateful to Jonathan Simon for helpful discussions. We gratefully acknowledge financial support by the Austrian Science Fund (FWF; SFB FoQuS Project No. F 4017 and DK CoQuS Project No. W 1210-N16) and the European Commission (IP SIQS, No. 600645). A.H. acknowledges financial support the Austrian Science Fund (FWF: Meitner Program Project M 1970).


\section*{Competing Interests}
\noindent The authors declare that they have no
competing financial interests.

\section*{Correspondence}
\noindent  Correspondence and requests for materials
should be addressed to J.V. (email: jvolz@ati.ac.at) and A.R. (email: arno.rauschenbeutel@ati.ac.at).

\newpage

\clearpage\newpage

\begin{figure}[tb]
	\centering
	\includegraphics[width=9cm]{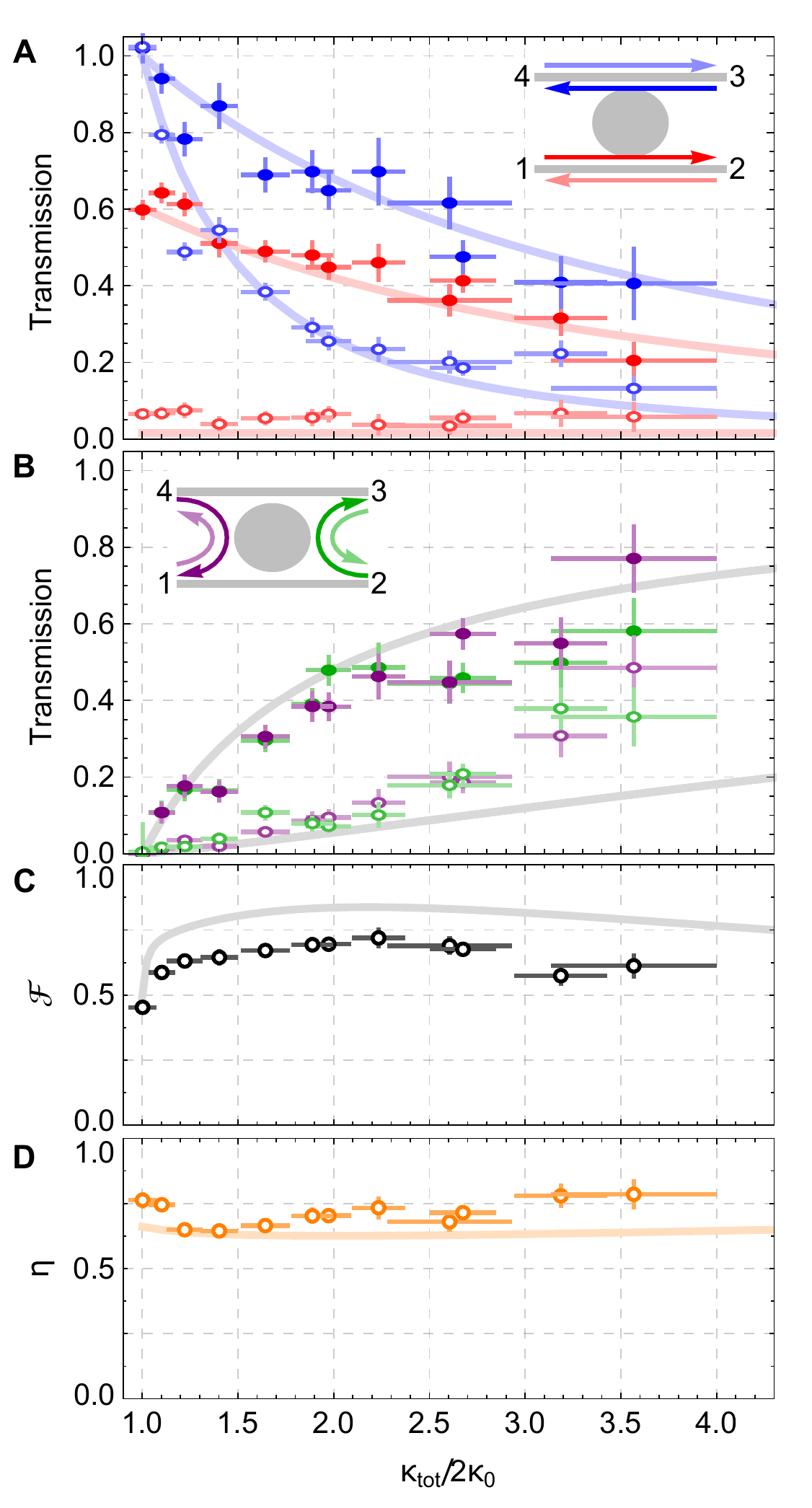}
	\caption{\textbf{Circulator performance.} (\textbf{A}) \& (\textbf{B}) Port-to-port transmissions as a function of the normalized field decay rate of the fiber-coupled resonator, $\kappa_{\rm tot}/2\kappa_0$, in the presence of an atom prepared in the $F=3$, $m_F=+3$ Zeeman state. Here, $\kappa_0=2\pi\times 5$~MHz is the intrinsic field decay rate of the resonator. The solid lines in both panels are the predictions of our theoretical model~\cite{supplementaries} with the atom-resonator coupling strength $g_{\rm ccw}=2\pi\times12$~MHz. (\textbf{C}) \& (\textbf{D}) Operation fidelity $\mathcal{F}$ and photon survival probability $\eta$ of the circulator, calculated from the data in \textbf{A} and \textbf{B}. The solid lines are the predictions of the above model for the same value of $g_{\rm ccw}$. The vertical error bars indicate $\pm 1\sigma$ statistical errors while the horizontal error bars represent an estimate of the variation of $\kappa_{\rm tot}$ due to drifts of the distances between the fiber couplers and the resonator during the corresponding measurement.}
	\label{fig:transmission_scan}
\end{figure}

\clearpage\newpage

\begin{figure}[tb]
	\centering
	\includegraphics[width=9cm]{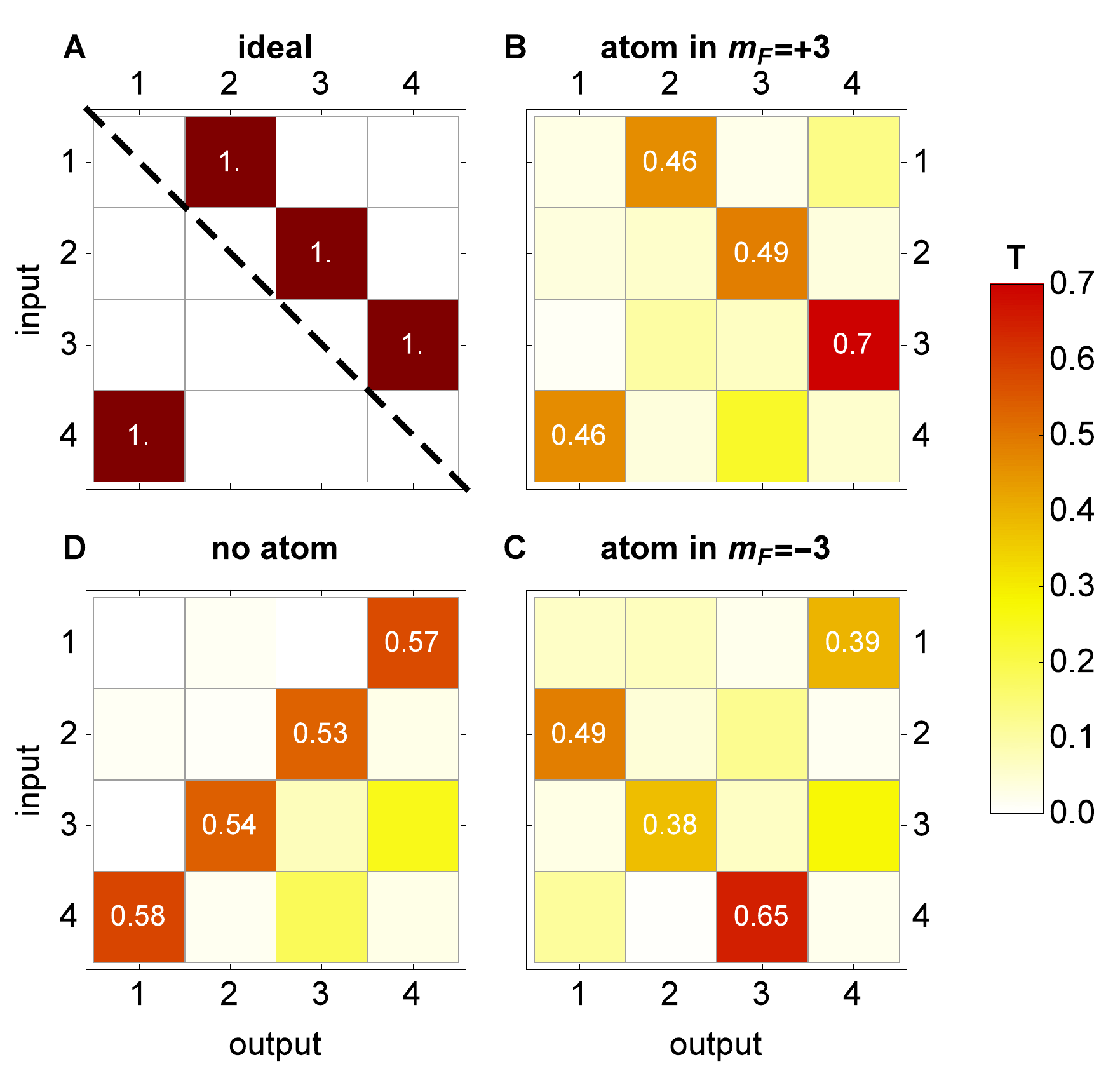}
	\caption{\textbf{Transmission matrices $\mathbf{(T_{i,j})}$.} The rows correspond to the input ports $i$ and the columns to the output ports $j$. (\textbf{A}) Transmission matrix for an ideal cirulator. The broken symmetry with respect to the dashed line indicates the nonreciprocal character of the device.  (\textbf{B}) Transmission matrix measured for the atom prepared in the $F=3$, $m_F=+3$ state. (\textbf{C}) Transmission matrix if the operation direction of the circulator is reversed by preparing the atom in the $F=3$, $m_F=-3$ state. For comparison, (\textbf{D}) shows the transmission matrix of the system measured without atom. Here, the symmetric matrix indicates reciprocal operation. For \textbf{B}, \textbf{C}, and \textbf{D} the fiber--resonator coupling was set to the optimal working point $\kappa_{\rm tot}/2\kappa_0=2.2$. In all panels, the four highest transmission values are given in the respective block.}
	\label{fig:transmission_matrix}
\end{figure}
 
\clearpage\newpage 

\clearpage\newpage

\section*{Materials and Methods}

\subsection*{Experimental procedure}\label{app:resonator} 
The experimental sequence starts with an atomic fountain delivering a cloud of laser-cooled $^{85}$Rb atoms to the resonator. In order to detect the presence of a single atom in the resonator mode in real time, we critically couple fiber \textit{a} to the resonator which is loaded with fiber $b$ ($\kappa_a=\kappa_0+\kappa_b$). We send a detection light field into port 1 that is resonant with the empty resonator mode and with the $D_2$ transition of $^{85}$Rb. When an atom enters the resonator mode, the interaction with the detection light field optically pumps the atom into the $\ket{F=3,m_F=+3}$ hyperfine ground state and the transmission through fiber $a$ increases by two orders of magnitude. Single-photon-counting modules (SPCM) that are connected to output port 2 record the transmission of the detection light field through fiber \textit{a} (port 1 $\rightarrow$ 2). Using a field programmable gate array-based real-time detection and control system, we react to the transmission increase within approximately 150 ns: the detection light is switched off and a probe light field, resonant with the $|F=3, m_F=+3\rangle\rightarrow |F'=4, m_F=+4\rangle$-transition, is sent into port \textit{i} (mean photon flux (8, 6, 11, 6) photons/$\mu$s) for 400 ns. Subsequently, a 1$\mu s$ long re-detection interval ensures that the atom is still coupled to the resonator mode. A full experimental cycle consists of four of the described sequences where the input port number $i$ for the probe light field is incremented consecutively ($i \in \{1,2,3,4\}$). 

We select the first 200~ns of the probing window for analysis since for longer probing times the 2 $\rightarrow$ 3 and 4 $\rightarrow$ 1 performance of the circulator is affected by optical pumping effects due to the finite coupling of the atom to the clockwise propagating resonator mode, $g_\text{\rm cw} \neq 0$. 
In future applications this optical pumping could always be counteracted by employing an external or an additional fiber-guided light field that permanently pumps the emitters towards the desired internal state. Choosing this light to be resonant to a different optical transition than the one used for the operation of the isolator would then enable continuous operation of the circulator. 

For the measurement of reversed operation direction of the circulator, the detection light field was sent from port 2 $\rightarrow$ 1 through fiber \textit{a}, thereby optically pumping the atom into the $\ket{F=3,m_F=-3}$ hyperfine ground state. In order to realize efficient optical pumping, a small bias magnetic field of 1.5~G is applied along the resonator axis for both cases.

\subsection*{Transmission measurements}
For the transmission measurements in Figs.~\ref{fig:transmission_scan}A and \ref{fig:transmission_scan}B, we scan $\kappa_\text{\rm tot}/2 \kappa_\text{0}$ by scanning the distance between fiber $b$ and the resonator surface. For each step, we subsequently position fiber $a$ such that $\kappa_a$ fulfills the critical coupling condition for the empty resonator which is loaded with fiber $b$. We determine the intrinsic field decay rate, $\kappa_\text{0}$, from the empty resonator line width without fiber $b$. For each scan point, the two fiber--resonator coupling rates, $\kappa_a$ and $\kappa_b$, are then inferred from the respective line width of the empty resonator which is loaded with both fibers, $a$ and $b$.

A direct measurement of the circulator transmissions is not possible because of the a priori unknown input and output losses imposed by the auxiliary fiber network that is used to feed the probe light into and out of the tapered fiber couplers as well as the a priori unknown efficiencies of the photo-detectors. We therefore devise the following measurement strategies for three different groups of input--output port combinations:

(i) The light is forward-transmitted through a given coupling fiber ($T_{1,2}$\,, $T_{2,1}$\,, $T_{3,4}$\,, $T_{4,3}$): We measure the corresponding output signal of the circulator including the auxiliary fiber network and normalize this value to the same output signal of the network when the WGM bottle microresonator is far detuned from the probe light frequency. This directly yields the corresponding transmissions which are plotted in Fig.~\ref{fig:transmission_scan}A.

(ii) The light is transferred from a given input port to the adjacent output port of the other coupling fiber ($T_{1,4}$\,, $T_{2,3}$\,, $T_{3,2}$\,, $T_{4,1}$): We measure the corresponding output signal of the circulator including the auxiliary fiber network and normalize this value to the same output signal of the network when no atom is coupled to the WGM bottle microresonator. We then multiply this normalized value by the theoretically predicted on-resonance transmission through the empty resonator, which is given by $T_{\text{cross}} = 1-2\kappa_0/\kappa_{\rm tot}$, cf. Eq.~(\ref{eq:22}) for $g=\Delta_{\text{rl}}=0$. This yields the corresponding transmissions  which are plotted in Fig.~\ref{fig:transmission_scan}B.

(iii) The light changes its propagation direction ($T_{i,i}$ and $T_{i,i+2}$ with $i\in\{1, 2, 3, 4\}$): We measure the corresponding output signal of the circulator including the auxiliary fiber network. From the design of the optical setup, we know that the output losses are approximately identical for ports 1 and 4. Taking advantage of this fact and using only the normalization measurements carried out for points (i) and (ii) above, we can then derive the corresponding transmissions. Their values do not exceed a few percent. For the optimal working point of the circulator of $\kappa_{\rm tot}/2\kappa_0=2.2$, they can be read from Fig.~\ref{fig:transmission_matrix} and are explicitly listed in section "Transmission matrices" below.

\subsection*{Modeling the circulator transmission}
\subsubsection*{Simplified model}
When light is sent into one of the input ports of fiber $a$ ($b$) of our system, it is partially transmitted through the fiber and partially cross-coupled to fiber $b$ ($a$) via the resonator. The respective output fields are given by 
\begin{equation}
 \begin{aligned}
 \left\langle \hat{a}^\text{trans}_{\text{out}}\right\rangle &=\left\langle \hat{a}_\text{in} \right\rangle-i \sqrt{2 \kappa_{a/b}}\left\langle \hat{a} \right\rangle \\
 \left\langle \hat{a}^\text{cross}_{\text{out}}\right\rangle &=-i \sqrt{2 \kappa_{b/a}}\left\langle \hat{a} \right\rangle  \; .
 \label{eq:aout1}
 \end{aligned}
\end{equation}
where $\left\langle \hat{a}_\text{in} \right\rangle$ is the amplitude of the input field and $\left\langle \hat{a} \right\rangle$ is the expectation value  of the photon annihilation operator for the resonator field. The latter is obtained by solving the master equation of the atom-resonator system 
\begin{equation}
 \frac{d \hat{\rho}}{dt} = -\frac{i}{\hbar} [\hat{H},\hat{\rho}] + \mathcal{L} \hat{\rho} \;,
 \label{eq:MasterEq}
\end{equation}
where $\hat H$ is the Hamiltonian of the system and $\mathcal{L}$ is the Lindblad superoperator~\cite{Carmichael1999}. The operation principle of the circulator can be best understood when making the approximation that the two counter-propagating resonator modes have almost perfect circular polarization, and thus are orthogonally polarized. We can then decompose the atomic $V$-type level system into two independent two-level systems, each of which couples to only one of the resonator modes with a direction-dependent coupling strength $g_{\rm cw/ccw}$. In this case, both directions of the circulator (forward and backward) can be described by the Jaynes-Cummings Hamiltonian which, in rotating wave approximation, is given by
\begin{equation}
 \hat{H}/\hbar=\Delta_{rl}\hat{a}^\dagger \hat{a}+\Delta_{al}\hat{\sigma}_{+}\hat{\sigma}_{-} +g(\hat{a}^\dagger \hat{\sigma}_{-}+\hat{a} \hat{\sigma}_+) \nonumber \\ + \sqrt{2 \kappa_{a/b}} (\epsilon \hat{a}-\epsilon^* \hat{a}^\dagger)
 \label{eq:Ham2Level}
\end{equation}
where $\Delta_\text{rl}$ ($\Delta_\text{al}$) is the resonator-light (atom-light) detuning, $\hat{\sigma}_+$ ($\hat{\sigma}_-$) the atomic excitation (deexcitation) operator, $\epsilon = \left\langle \hat{a}_\text{in} \right\rangle$ and the last term describes the pumping of the resonator via fiber $a$ or $b$, respectively. The corresponding Lindblad superoperator is given by 
\begin{equation}
 \mathcal{L}=\kappa_\text{\rm tot} ( 2\hat a\hat{\rho} \hat a^\dagger-\hat a^\dagger  \hat a \hat{\rho}-\hat{\rho}  \hat a^\dagger \hat a) + \gamma ( 2\hat \sigma_-\hat{\rho} \hat \sigma_+ -\hat \sigma_+  \sigma_- \hat{\rho}-\hat{\rho}  \hat \sigma_+ \sigma_-) 
 \label{eq:Lind2Level}
\end{equation}
where $\kappa_\text{\rm tot}=\kappa_0+\kappa_a+\kappa_b$ is the total field decay rate of the fiber-coupled empty resonator.
In the low saturation limit, this model yields the power transmissions for all input-output port constellations: 
\begin{equation}
 T^\text{trans}_\text{cw/ccw} = \left|\frac{\left\langle a^\text{trans}_{\text{out}}\right\rangle}{\left\langle a_\text{in}\right\rangle}\right|^2 =\frac{|\Gamma_\text{cw/ccw} +i \Delta_\text{rl}+\kappa_\text{i}+\kappa_\text{b(a)}-\kappa_\text{a(b)}|^2}{|\Gamma_\text{cw/ccw} +i \Delta_\text{rl}+\kappa_\text{\rm tot}|^2}
 \label{eq:12}
\end{equation}
\begin{equation}
 T^\text{cross}_\text{cw/ccw} =  \left|\frac{\left\langle \hat{a}^\text{cross}_{\text{out}}\right\rangle}{\left\langle \hat{a}_\text{in}\right\rangle}\right|^2 = \frac{4 \kappa_\text{a} \kappa_\text{b}}{|\Gamma_\text{cw/ccw} +i \Delta_\text{rl}+\kappa_\text{\rm tot}|^2}
 \label{eq:22}
\end{equation}
where we introduced the atom-induced field decay rates
\begin{equation}
 \Gamma_\text{cw/ccw} = \frac{g_\text{cw/ccw}^2}{\gamma+i \Delta_\text{al}} \; .
 \label{eq:3}
\end{equation}
$T^\text{trans}_\text{cw/ccw}$ describes the transmission through fiber $a$ ($b$) and $T^\text{cross}_\text{cw/ccw}$ the cross-coupling from both fiber $a$ to fiber $b$ and fiber $b$ to fiber $a$ via the resonator. In this model, the ratio $g_\text{\rm ccw}/g_\text{\rm cw}$ is determined by the difference of the transition strengths of the two independent two-level systems. \\

\subsubsection*{Coupling to both resonator modes}
In the previous model, we made the assumption of perfect chiral coupling, i.e., each resonator mode couples exclusively to a single atomic transition. However, for our experimental system, this situation is not fully realized as the evanescent field of the resonator is not fully circularly polarized. In order to describe this situation and to obtain an analytical solution, we model our system as a two-level atom with a circular dipole transition (here $\sigma^+$), i.e., we neglect the coupling of the resonator fields to the weaker transition in the atom. Using this approximation, the Hamiltonian is given by 
\begin{equation}
\hat{H}/\hbar=\Delta_{rl}\hat{a}^\dagger \hat{a}+\Delta_{al}\hat{\sigma}_{+}\hat{\sigma}_{-} +\alpha g(\hat{a}^\dagger \hat{\sigma}_{-}+\hat{a} \hat{\sigma}_+) +\beta g(\hat{b}^\dagger \hat{\sigma}_{-}+\hat{b} \hat{\sigma}_+) \nonumber \\ + \sqrt{2 \kappa_{a/b}} (\epsilon \hat{a}-\epsilon^* \hat{a}^\dagger)
\end{equation}
and the corresponding Lindblad superoperator reads
\begin{equation}
\mathcal{L}=\kappa_\text{\rm tot} ( 2\hat a\hat{\rho} \hat a^\dagger-\hat a^\dagger  \hat a \hat{\rho}-\hat{\rho}  \hat a^\dagger \hat a)+\kappa_\text{\rm tot} ( 2\hat b\hat{\rho} \hat b^\dagger-\hat b^\dagger  \hat b \hat{\rho}-\hat{\rho}  \hat b^\dagger \hat b) + \gamma ( 2\hat \sigma_-\hat{\rho} \hat \sigma_+-\hat \sigma_+  \sigma_- \hat{\rho}-\hat{\rho}  \hat \sigma_+ \sigma_-) .
\end{equation}
Here, $\hat a$ and $\hat b$ ($\hat a^\dagger$ and $\hat b^\dagger$) are the annihilation (creation) operators for a photon in the ccw and cw resonator mode, respectively, and 
$\alpha=|\textbf{E}_{\rm ccw} \cdot \textbf{e}_{\sigma^+}|/|\textbf{E}_{\rm ccw}|$ ($\beta=|\textbf{E}_{\rm cw} \cdot \textbf{e}_{\sigma^+}|/|\textbf{E}_{\rm cw}|$) is the overlap of the evanescent field $\textbf{E}_{\rm ccw}$  ($\textbf{E}_{\rm cw}$) of the ccw (cw) mode with the $\sigma^+$-polarization, $\textbf{e}_{\sigma^+}$, of the atomic transition. This model gives an accurate description for most physical situations because, typically, the coupling strength to the weak atomic transition is significantly smaller than the residual coupling to the strong atomic transition due to imperfect circular polarization. From this model, we obtain the theoretical transmission curves depicted in Fig. \ref{fig:transmission_scan}, taking into account the actual polarization properties of our resonator modes ($\alpha=\sqrt{0.97}$, $\beta=\sqrt{0.03}$). Solving this model, one obtains analytical expressions for the photon survival probability in forward and backward direction:
\begin{eqnarray}
\eta_\text{fw}&=&    1-\frac{2 (\kappa_\text{tot}-\kappa_0)  \left(\gamma ^2 \kappa_0 \kappa_\text{tot}^2+g^4
	\kappa_0 \alpha ^2 +  g^2 \gamma\kappa_\text{tot}
	\left(\alpha ^2(2\kappa_0-\kappa_\text{tot})+2 \kappa_\text{tot}\right)\right)}{\kappa_\text{tot}^2 \left(\gamma \kappa_\text{tot}+g^2\right)^2} \\
\eta_\text{bw}&=&    \frac{1}{\kappa_\text{tot}^2\left(\gamma \kappa_\text{tot}+g^2\right)^2}\Big[
	(\gamma ^2(\kappa_\text{tot}-\kappa_0)^2+\left(\gamma
	\kappa_0+g^2\right)^2)\kappa_\text{tot}^2 \nonumber\\
&&	\hspace{3cm}-2g^2 (\kappa_\text{tot}-\kappa_0) 
	(1-\alpha^2) \left(\gamma\kappa_\text{tot}  \left( 2\kappa_0-\kappa_\text{tot}\right)+g^2\kappa_0\right)\Big]\,.
\end{eqnarray}
From this, we obtain the process fidelity and photon survival probability of the circulator
\begin{eqnarray}
\mathcal{F}& = &\frac{ 
	\frac{1}{\eta_\text{bw}} \left(\kappa _{\text{tot}}-\kappa _0\right)^2 \left(\alpha ^2 g^2+\gamma  \kappa _{\text{tot}}\right)^2 +
	\frac{1}{\eta_\text{fw}} \left(g^2 \left(\kappa_0+\alpha^2\left(\kappa _{\text{tot}}-\kappa_0\right)\right)+\gamma\kappa_0 \kappa _{\text{tot}}\right)^2} 
	{2 \kappa _{\text{tot}}^2 \left(g^2+\gamma  \kappa _{\text{tot}}\right)^2}\\
\eta &=&\frac{\eta_\text{fw}+\eta_\text{bw}}{2}= 1-(\kappa_{\rm 	tot}-\kappa_0)\frac{2\gamma^2\kappa_{\rm tot}^2\kappa_0+g^4\kappa_0+\gamma g^2\kappa_{\rm tot}(\kappa_{\rm tot}+2\kappa_0)    }
{\kappa_{\rm tot}^2(\gamma\kappa_{\rm tot}+g^2)^2}\,.
\end{eqnarray}
Here, we made the assumption that both fibers are equally well coupled to the resonator, i.e., $\kappa_a=\kappa_b=\kappa$. Interestingly, the photon survival probability is independent of the polarization overlap $\alpha$. For parameters achievable with state-of-the-art WGM resonators~\cite{Poellinger2009,Junge13} ($\kappa_0=2\pi\times0.5$ MHz, $g=2\pi\times 30$ MHz, $\alpha=\sqrt{0.97}$) one reaches $\eta=0.94$ and $\mathcal{F}=0.94$ for an optimum fiber-resonator coupling rate $\kappa=2\pi\times 7.5$ MHz. For this case, the process fidelity is mainly limited by the non-unit overlap of the evanescent field with $\sigma^+$ polarization ($\alpha^2<1$).

\subsection*{Second order correlation functions}
\begin{figure}[t]
	\centering
	\includegraphics[width=12cm]{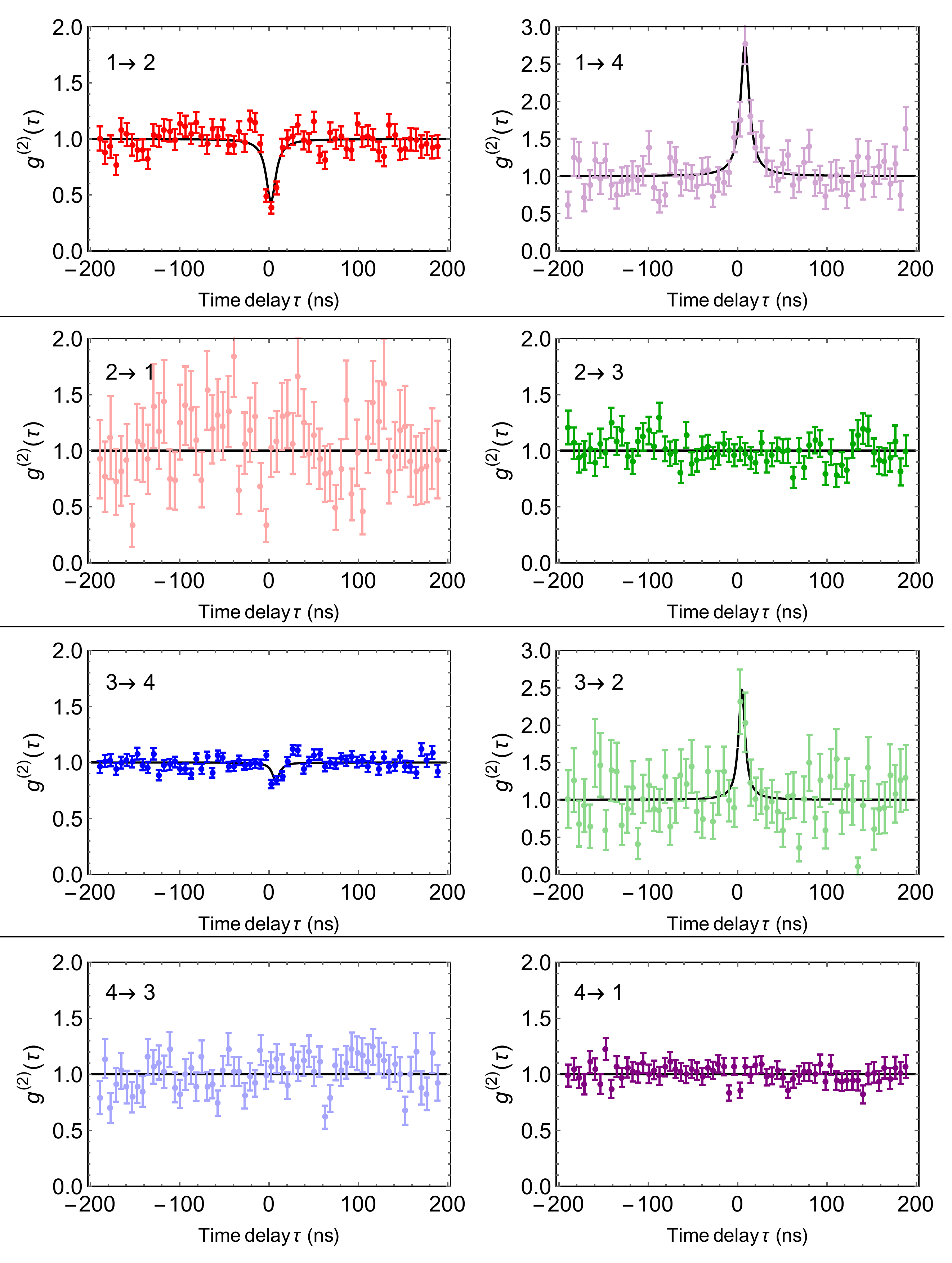}
	\caption{Second-order correlation function measurements in all relevant output ports for all four input directions including a Lorentzian fit as a guide to the eye. }
	\label{fig:g2_all}
\end{figure}

In Fig.~\ref{fig:g2_all} we plot the full set of second-order correlation functions. For those pairs of ports not shown, i.e., $i\rightarrow i$ and $i\rightarrow i+2$), the total number of transmitted photons is too small to derive a meaningful second-order correlation function. 

\subsection*{Transmission matrices}\label{subsection:Transmission_matrices}
The transmission values $T_{i,j}$  are presented with their error bars indicating the $\pm 1\sigma$ statistical error.  The rows correspond to the respective input ports and the columns to the respective output ports. The transmissions are measured for $\kappa_{\rm tot}/2\kappa_0=2.2$. $\text{T}^{m_F=+3}$ is the transmission matrix for the case where the atom is prepared in the $F=3$, $m_F=+3$ state, $\text{T}^{m_F=-3}$ for the atom prepared in the $F=3$, $m_F=-3$ state and $\text{T}^{\rm no~atom}$ for the case where the atom is not present.
  
\begin{equation}
\text{T}^{m_F=+3}=
\left(
\begin{array}{cccc}
 0.03\pm 0.015 & 0.46\pm 0.044 & 0.024\pm 0.011 & 0.133\pm 0.03 \\
 0.037\pm 0.022 & 0.057\pm 0.021 & 0.486\pm 0.059 & 0.038\pm 0.022 \\
 0.011\pm 0.011 & 0.101\pm 0.028 & 0.068\pm 0.022 & 0.698\pm 0.083 \\
 0.463\pm 0.055 & 0.039\pm 0.014 & 0.234\pm 0.027 & 0.055\pm 0.019 \\
\end{array}
\right)
\end{equation}

\begin{equation}
\text{T}^{m_F=-3}=
\left(
\begin{array}{cccc}
 0.063\pm 0.01 & 0.072\pm 0.013 & 0.021\pm 0.007 & 0.394\pm 0.025 \\
 0.487\pm 0.028 & 0.045\pm 0.01 & 0.122\pm 0.017 & 0.016\pm 0.005 \\
 0.029\pm 0.007 & 0.379\pm 0.029 & 0.066\pm 0.012 & 0.274\pm 0.021 \\
 0.108\pm 0.011 & 0.005\pm 0.001 & 0.647\pm 0.029 & 0.02\pm 0.005 \\
\end{array}
\right)
 \end{equation}

\begin{equation}
\text{T}^{\rm no~atom}=
\left(
\begin{array}{cccc}
 0.\pm 0. & 0.014\pm 0.008 & 0.\pm 0. & 0.572\pm 0.061 \\
 0.012\pm 0.012 & 0.008\pm 0.008 & 0.533\pm 0.062 & 0.025\pm 0.018 \\
 0.\pm 0. & 0.539\pm 0.063 & 0.075\pm 0.023 & 0.252\pm 0.052 \\
 0.583\pm 0.06 & 0.016\pm 0.009 & 0.183\pm 0.025 & 0.027\pm 0.014 \\
\end{array}
\right)
\end{equation}


\begin{thebibliography}{10}
\bibitem{supplementaries}
Detailed materials and methods can be found on science online.
\end{thebibliography}
\end{document}